\documentclass[12pt,oneside,notitlepage,abstracton,a4paper]{scrartcl}
\usepackage{epsfig,scrpage2,graphicx}
\usepackage{lineno}

\lehead{EUROTeV-Report-2006-022}
\lohead{EUROTeV-Report-2006-022}

\titlehead{EUROTeV-Report-2006-022}
\subject{\includegraphics[bb=0 0 142 91]{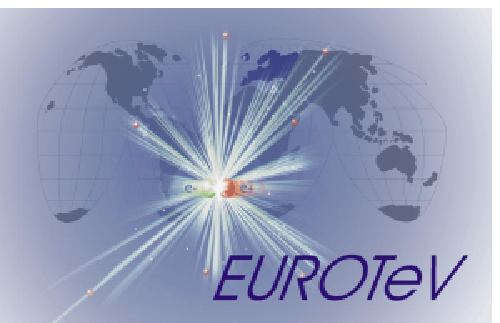}}

\title{\Large Improved final doublet designs for the ILC baseline small crossing angle scheme} 
\author{\normalsize R. Appleby\thanks{r.b.appleby@dl.ac.uk, The Cockcroft Institute and the University of Manchester, Oxford Road, 
Manchester, M13 9PL, UK},
P. Bambade\thanks{Laboratoire de l'Acc\'{e}lerat\'{e}ur Lin\'{e}aire, CNRS/IN2P3 and Universit\'{e} Paris Sud 11, B\^{a}timent 200, B.P. 34, F-91898 Orsay, Cedex, France}}

\date{\normalsize \today}

\begin{document}

\maketitle

\begin{abstract}
The ILC baseline consists of two interaction regions, one with a 20mrad crossing angle and the other
with a 2mrad crossing angle. It is known that the outgoing beam losses in the final doublet and
subsequent extraction line are larger in the 2mrad than in the 20mrad layout. In this work, we exploit
NbTi and Nb$_3$Sn superconducting magnet technologies to redesign and optimise
the final doublet, with the aim of providing satisfactory outgoing disrupted beam
power losses in this region. We present three new final doublet layouts, specifically optimised
for the 500 GeV and the 1 TeV machines.
\end{abstract}

\newpage

\section{INTRODUCTION}


The current baseline configuration of the International Linear Collider~\cite{ilc} consists of
two interaction regions (IRs) and two detectors, with one IR having a large beam crossing angle and
one having a small beam crossing angle.
The large crossing angle layout has an angle of 20mrad
between the beams, and is derived from the large crossing angle design
of the Next Linear Collider~\cite{nlc}. Its optical properties are well understood, and the extraction
line successfully transports the disrupted beam to the dump, for a wide range of proposed beam parameter sets~\cite{schulte}. Furthermore, the design includes downstream diagnostics for energy and polarisation
measurement. The 2mrad interaction region layout presents more technical challenges, mainly
resulting from transporting the outgoing disrupted beam off-axis in the final doublet. This design is more recent and the first complete
optics was presented at Snowmass 2005~\cite{2mradsnowmass}. This layout was developed
for 1 TeV using NbTi superconducting final doublet magnets and included downstream
energy and polarimetry measurement chicanes.

The current ILC parameter space is documented in~\cite{ilcparas} and the key parameters are listed in Table~\ref{tabparasets}. These consist of sets of representative beam parameters, designed to assure 
flexibility in case of unexpected practical limitations in achieving some of the machine goals. The most challenging
of these sets is known as the High Luminosity set, which aims to maximise the instantaneous luminosity by more strongly focusing the beams at the IP, thereby 
significantly enhancing disruption effects. A necessary criterion for the performance of the extraction line design is its ability to
accommodate the broadest possible range of beam parameters. The 20mrad scheme is successful for the majority of parameter sets,
but the current (baseline) 2mrad design shows unacceptable beam power loss, both in the
final doublet and in the extraction line, for some of the beam parameters considered. 

In this work, we exploit the limits of current magnet technology as well as technology still under development
to optimise the 2mrad final doublet layout. We present three new final doublet designs, one optimised for
the baseline energy of 500 GeV and two for the upgrade energy of 1 TeV. We optimise the
magnet parameters with respect to power deposition from charged beam and radiative Bhabha particle losses, with the goal of
accommodating all ILC beam parameter sets. We find that this is possible by using NbTi technology for the 500 GeV machine, and
by using new Nb$_3$Sn technology for the 1 TeV upgrade. However, an increase in pole tip field, somewhat beyond the currently
planned limits for Nb$_3$Sn technology, would be required to achieve power losses small enough to avoid quenching of the final doublet
magnets for the case of the Low Power and High Luminosity beam parameters at 1 TeV. We include a full set of magnet and optical
parameters for evaluation by magnet designers. 

In Section~\ref{secmags} we discuss the available magnet technologies and the constraints
we place on magnet parameters. We describe our final doublet optimisation
procedure in Section~\ref{secopt} and present the resulting magnet parameters for each case. The
beam transport and resulting losses with our new doublets are presented in 
Section~\ref{secpl}, and we draw our conclusions in Section~\ref{secconc}. We find that the
proposed new final doublet layouts have considerably better beam transport properties
than the current ones. 

\begin{table}[htb]
\vspace{-0mm}
\begin{center}
\caption{Key parameters of the beam parameter sets used in this work.}
\vspace{0mm}
\begin{tabular}{|l|c|c|c|c|c|}
\hline
Set		& Energy  & N$_b$  & I$_{av}$ [A] & $\delta_B$ & $\mathcal{L}$ [m$^{-2}$ s$^{-1}$] \\ \hline \hline
Nominal 	& 500 GeV & 2820 &  0.0104 & 0.022 & 2.03E38 \\ \hline	
Low Power	& 500 GeV & 1330 &  0.0069 & 0.057 & 2.05E38 \\ \hline
High Lum.	& 500 GeV & 2820 &  0.0104 & 0.070 & 4.92E38 \\ \hline
Nominal 	& 1 TeV   & 2820 &  0.0104 & 0.050 & 2.82E38 \\ \hline	
Low Power	& 1 TeV   & 1330 &  0.0069 & 0.134 & 2.92E38  \\ \hline
High Lum.	& 1 TeV   & 2820 &  0.0104 & 0.178 & 7.88E38  \\ \hline
\end{tabular}
\label{tabparasets}
\end{center}
\end{table}

\section{MAGNET PARAMETERS}
\label{secmags}

\subsection{Current baseline design}

The final doublet in the 2mrad scheme is based a superconductive large bore QD0 magnet and on a normal conducting smaller aperture QF1 magnet. The two nearby 
sextupoles for local chromaticity correction, SD0 and SF1, are also superconductive large bore magnets. In the current design, the choice of superconductive technology for 
QD0 is NbTi and the maximum pole tip field which is assumed to be achievable is 5.6 T. In this work we always take l$^*$ to be 4.5m and the
QD-QF distance to be 9m. The outgoing beam passes through the coil pocket of QF1, where 
the field is modelled by a standard multipole expansion. The outer coil radius, which is
available for beam passage, is 115mm. In this design, the doublet was optimised for the TeV machine, and the
500 GeV layout was obtained by scaling down the fields.

\subsection{Maximum pole tip field for NbTi}

For the NbTi magnets~\cite{saclay} considered in this work, we assume a maximum pole tip field of 6.28 T. This is
around the maximum field that current coil designs can produce, and includes the field-reducing effects of the 
external solenoid and a safety magin~\cite{brettparker}. This corresponds to a gradient of 180 T m$^{\mathrm{-1}}$ in a 
magnet of radial aperture 35mm. The sextupoles are required to be
large bore superconducting magnets, to allow the transport of the outgoing beam. We follow
the technology choice of the current design, and adopt a maximum field of 4.04T in SD0 and
SF1 (with radial apertures of 88mm and 112mm, respectively)\footnote{The maximum field in a sextupole probably has some dependance on its aperture. 
This point should be checked with experts, to assess the feasibility of our proposed magnets.}.

\subsection{Expected maximum pole tip field for Nb$_3$Sn}

We also propose to use Nb$_3$Sn-based technology~\cite{saclay} for QD0 at 1 TeV. The higher gradient permitted
by this magnet will improve the beam transport properties of the doublet, by allowing a
more compact design. These magnets have yet to be proven, but in the timescales of the
ILC TeV upgrade, should hopefully be readily 
available\footnote{The development of superconductive magnets based on Nb$_3$Sn technology is also being strongly 
stimulated by the LHC luminosity upgrade programme, which also requires higher gradients for its low-beta insertion quadrupoles.}. 
Of course, such an upgrade would require replacing the final doublet, but this is rather minor on the scale of the 1 TeV energy upgrade.
Furthermore the geometry of the extraction line would need to 
be maintained. The maximum pole tip field which we assume will be achievable with Nb$_3$Sn is 8.8T (taking into account the presence of a solenoid background field). 
This corresponds to a 250 T m$^-1$ gradient for a 35mm radial aperture. 

\subsection{Outline of doublet designs}

In this work, we propose three new doublet layouts. We shall optimise the layout
for the 500 GeV machine, using NbTi for QD0 at a maximum pole tip field of 6.28 T. The beam transport
properties of such a layout will be considerably better that the scaled down version
of the 1 TeV existing baseline, because the magnet length can be reduced. The sextupoles shall be optimised versions of the current
baseline. We shall then calculate the layouts for the TeV machine, firstly
using NbTi for QD0 as the design for 500 GeV, and secondly with Nb$_3$Sn. The resulting doublets are intended to replace the
present design.

The current baseline magnets are
summarised in Table~\ref{tabmpbl}. In Table~\ref{tabmp} we describe our proposed magnet technology limits for this work.

\begin{table}[htb]
\vspace{-0mm}
\begin{center}
\caption{Quadrupole and sextupole parameters at $\sqrt{s}=$1 TeV for the current
baseline design.}
\vspace{0mm}
\begin{tabular}{|l|c|c|c|c|c|}
\hline
Magnet	& Length  & Strength & radial aperture & gradient & B$^{\mathrm{PT}}$ \\ \hline \hline
QD0 	& 2.5m & -0.0958 m$^{-1}$    & 35mm		  & 159.4 T m$^{\mathrm{-1}}$  & 5.58 T \\ \hline	
SD0	& 3.8m	& 0.6254 m$^{-2}$    & 88mm		  & -	     & 4.04 T \\ \hline
QF1	& 2.0m	& 0.0407 m$^{-1}$    & 10mm		  & 67.8 T m$^{\mathrm{-1}}$   & 0.68 T \\ \hline
SF1	& 3.8m	& -0.2039 m$^{-2}$   & 112mm		  & -	     & 2.13 T \\ \hline
\end{tabular}
\label{tabmpbl}
\end{center}
\end{table}

\begin{table}[htb]
\vspace{-0mm}
\begin{center}
\caption{The limiting parameters of the final doublet magnets. Note that these maximum fields include the reductions expected from the solenoid background field at the location of the magnets.}
\vspace{0mm}
\begin{tabular}{|l|c|c|}
\hline
Magnet	& Type & Maximum field \\ \hline \hline
QD0 (NbTi) 	& SC   & 6.28 T	       \\ \hline	
QD0 (Nb$_3$Sn) 	& SC   & 8.75 T	       \\ \hline
SD0		& SC	& 4.04 T (r=88mm)	   \\ \hline
QF1		& NC	& 1.02 T  \\ \hline
SF1		& SC	& 2.13 T (r=112mm)  	\\ \hline
\end{tabular}
\label{tabmp}
\end{center}
\end{table}

\subsection{Assumed power deposition tolerances}
\label{secpowerdepos}

Superconductive magnets will quench if too much beam power is deposited in
them, and both integrated power and localised  power density tolerances have
been specified by the CERN group for the LHC quadrupoles used in this work. The
maximum  integrated power which can be tolerated per unit length is in the
range 5-10 W m$^{\mathrm{-1}}$. However, for a few specific magnets like those in the final
doublet, this tolerance can presumably be relaxed a bit if more cryogenic
power can be used. The highest  localised power density which can be
accepted is 0.5 mW g$^{\mathrm{-1}}$. This tolerance is important to consider in this work 
because the outgoing beam is transported off-axis
through several of the final doublet magnets, and so particles at the lower end
of the energy distribution can get over-focused onto the edge of the magnet
aperture on one side. Since these particles are essentially concentrated
in a plane (except for their vertical angular spread), the local power
density can become high. We rely on detailed computations performed at RHUL
with the GEANT-4 based BDSIM simulation~\cite{privatejohn}, following similar calculations 
at SLAC~\cite{privatelew}. These calculations have shown that for about 1
W incident beam power on the edge of QD0, a maximum power density of about
4 mW g$^{\mathrm{-1}}$ can be reached inside the NbTi coils. If a 3 mm Tungsten liner
(corresponding to about 1  radiation length) is used to thicken the beam
pipe and spread out the showers at the location of the loss, the maximum
power density can be brought down to about 0.3 mW g$^{\mathrm{-1}}$ in the same conditions.
Based on these numbers, we use 1 W as the  maximum value for the power
deposited on the edge of QD0. This 1 W $\leftrightarrow$ 0.3 mW g$^{\mathrm{-1}}$ equivalence, found
between total incident power and power density in QD0, will have to be
checked for other magnet configurations and beam parameters, especially in
the case of the sextupoles.

\section{FINAL DOUBLET OPTIMISATION}
\label{secopt}

In this section, we shall discuss our optimisation of the final doublet layout.
We shall start with
the optimisation of QD0. Several studies of the power deposition into QD0 have shown that the charged
beam loss and the loss from radiative Bhabhas exceed the quench limits of the coils, for some of the proposed beam parameters. Therefore,
the goal of our optimisation is to reduce these losses to acceptable levels, for all beam parameters. 
Note that we work in terms of total incident power, and consider the conversion to localised losses separately (see Section~\ref{secpowerdepos}). 
The optimisation procedure described in the following is applied to all three final doublet designs outlined in Section~\ref{secmags}.

The underlaying principle is, in the first instance, to make the magnet as
short as possible within the constraints of the available pole tip field. Hence we begin by
increasing the field of QD0 to the maximum possible value, and decrease the length in proportion.

We then perform a scan of the aperture and pole tip field, adjusting the length to maintain the overall focusing strength, and using the total power deposition from the different beam parameter sets as figure of merit. 
The magnitude of the beamstrahlung energy loss for the Low Power and High Luminosity beam parameters it such that these sets dominate the optimisation. Both the effects from the charged beam, for vertical offsets at the interaction point maximising the beamstrahlung losses, and the combined effects of radiative Bhabhas and the charged beam (collided in this case without a significant vertical offset) are considered. 
We generally find a mimimum in the total power deposition which is 
very similar for all parameters sets, as the dependance of the total power versus aperture and length found from the scanning is rather shallow. 
This procedure allows a power-optimised QD0 length, aperture and pole tip field, with the constraints of maintaining the constant integrated focusing strength needed to achieve the required 
demagnification and keeping the pole tip below a given assumption on the maximum value achievable. Note that the former constraint 
maintains the demagnification
of the final telescope, and also maintains the integrated strength of QF1. To ensure that the
mimimum we reach in the 3-dimensinal space of the scan is a true and  global minimum, we
begin with a variety of initial magnet conditions (i.e. origin in our space). We find that 
we always converge to the same point in all cases. We perform this optimisation
for the three cases we consider in this work and the result is a set of three QD0-QF1 systems.

Once QD0 has been optimised, we can fit for the local chromaticity correction sextupoles. For these sextupoles, located in the immediate vicinity of the final doublet quadrupoles, 
we assume the same maximum pole tip fields as in the baseline. The chromaticity is corrected, and the required sextupole strengths calculated, by
using a simplified model of the system, consisting of the final doublet region and two upstream bends to generate the required dispersion. The bend strengths
are fitted to give $\eta_x=0$ and $\eta_x'=-10\mathrm{mrad}$ at the interaction point. Note that the latter value is chosen to give acceptable sextupole strengths, whilst maintaining the beam 
angular divergence close to its intrinsic value. While a full treatment of
the final focus chromaticity correction includes the condition of zeroing
the $\mathrm{T}_{126}$ and $\mathrm{T}_{346}$ matrix element across the full system, it has been shown
that, to a good approximation, sextupole strengths consistent with those in
such a full treatment can be obtained by zeroing the $\mathrm{T}_{166}$ and $\mathrm{T}_{346}$  matrix
elements across the simplified system~\cite{uspaslc}. We apply such a
procedure to compute the sextupole strengths in this work. The inter-magnet distances chosen are shown
in Table~\ref{tabdrifts}. In this work, the optics code MAD~\cite{mad} was used to compute the chromaticity correction.

The length of the optimised sextupoles was chosen to give a reasonable starting point for the beam transport. The sextupole lengths are hence not fully optimised, and 
require some further study. To optimise the sextupole apertures, a fit is performed, within the maximal value assumed for the pole tip fields, and requiring no loss when transporting a Nominal, Low Power or High Luminosity beam. The
inter-magnet spacings were optimised for beam transport.
The technological constraints limit the maximum aperture,
and we fit the aperture by requiring no loss when transporting a nominal, low power or high luminosity 
beam. The available aperture is computed from the quadratic rise of the field in the beampipe
of the sextupole. Note that we consider both the radiative Bhabha contribution to the power deposition and the effect of a vertically
offset beam. The requirement of small enough loss (1 W in total) is very stringent, particularly in the case of the High Luminosity offset beams, and results in large sextupole apertures. 
These could be considerably reduced by relaxing the beam transport requirement, for instance if one would avoid beam parameter optimisations with very large beamstrahlung energy losses.

The results of the optimisation can be found in table~\ref{tabmpnbti500} for NbTi at 500 GeV, in table~\ref{tabmpnbti1000} for NbTi at
1 TeV and in table~\ref{tabmpnb3sn1000} for Nb$_3$Sn at 1 TeV. These tables show the magnet parameters of the two quadrupoles and two sextupoles
making up the final doublet layout. 

\begin{table}[htb]
\vspace{-0mm}
\begin{center}
\caption{The inter-magnet drift spaces used in the final doublet. In all the layouts, l$^*$ is
taken to be 4.5m, and the total QD-QF distance is always 9m.}
\vspace{0mm}
\begin{tabular}{|l|c|c|c|}
\hline
&		NbTi, 500 GeV	&	NbTi, 1 TeV 	& 	Nb$_3$Sn, 1 TeV \\ \hline\hline
QD0-SD0	&	0.8m		&	0.3m		&	0.3m \\ \hline
SD0-QF1	&	5.7m		&	4.9m		&	4.9m \\ \hline
QF1-SF1	&	0.5m		&	0.3m		&	0.3m \\ \hline
\end{tabular}
\label{tabdrifts}
\end{center}
\end{table}

\begin{table}[htb]
\vspace{-0mm}
\begin{center}
\caption{Quadrupole and sextupole parameters at $\sqrt{s}=$500 GeV for NbTi.}
\vspace{0mm}
\begin{tabular}{|l|c|c|c|c|c|c|}
\hline
Magnet	& Length  & Strength & radial aperture & gradient & B$^{\mathrm{PT}}$  \\ \hline \hline
QD0 	& 1.23m & -0.1940 m$^{-1}$    & 39mm		  & 161.6 T m$^{\mathrm{-1}}$ & 6.30 T    	\\ \hline	
SD0	& 2.5m	& 1.1166 m$^{-2}$    & 76mm		  & -	     & 2.69 T 	   \\ \hline
QF1	& 1.0m	& 0.0815 m$^{-1}$    & 15mm		  & 67.9 T m$^{\mathrm{-1}}$   & 1.02 T   \\ \hline
SF1	& 2.5m	& -0.2731 m$^{-2}$   & 151mm		  & -	     & 2.59T   	   	\\ \hline
\end{tabular}
\label{tabmpnbti500}
\end{center}
\end{table}

\begin{table}[htb]
\vspace{-0mm}
\begin{center}
\caption{Quadrupole and sextupole parameters at $\sqrt{s}=$1 TeV for NbTi.}
\vspace{0mm}
\begin{tabular}{|l|c|c|c|c|c|c|}
\hline
Magnet	& Length  & Strength & radial aperture & gradient & B$^{\mathrm{PT}}$  \\ \hline \hline
QD0 	& 2.47m & -0.0970 m$^{-1}$    & 39mm		  & 161.7 T m$^{\mathrm{-1}}$  & 6.31 T    \\ \hline	
SD0	& 3.8m	& 0.5993 m$^{-2}$    &   87mm		  & -	     &    3.78 T    	   \\ \hline
QF1	& 2.0m	& 0.0407 m$^{-1}$    & 15mm		  & 67.8 T m$^{\mathrm{-1}}$   & 1.02 T  \\ \hline
SF1	& 3.8m	& -0.1644 m$^{-2}$   & 146mm		  & -	     &	  2.92 T       	\\ \hline
\end{tabular}
\label{tabmpnbti1000}
\end{center}
\end{table}

\begin{table}[htb]
\vspace{-0mm}
\begin{center}
\caption{Quadrupole and sextupole parameters at $\sqrt{s}=$1 TeV for Nb$_3$Sn.}
\vspace{0mm}
\begin{tabular}{|l|c|c|c|c|c|c|}
\hline
Magnet	& Length  & Strength & radial aperture & gradient & B$^{\mathrm{PT}}$  \\ \hline \hline
QD0 	& 2.0m & -0.12 m$^{-1}$   & 44mm		  & 200 T m$^{\mathrm{-1}}$  & 8.80 T 	\\ \hline	
SD0	& 3.8m	& 0.6454 m$^{-2}$    & 95.1mm  		  & -	     & 4.87 T         \\ \hline
QF1	& 2.0m	& 0.0407 m$^{-1}$    & 15mm		  & 67.8 T m$^{\mathrm{-1}}$   & 1.02 T \\ \hline
SF1	& 3.8m	& -0.1689 m$^{-2}$   & 163mm		  & -	     &	3.74 T    	   	\\ \hline
\end{tabular}
\label{tabmpnb3sn1000}
\end{center}
\end{table}

\section{PARTICLE LOSSES IN THE OPTIMISED FINAL DOUBLET}
\label{secpl}

In this section, power depositions from particle losses are computed for the three optimised final doublets described in Section~\ref{secmags}. For completeness
we include the power losses for the current baseline layout, both at 500 GeV and 1 TeV. The disrupted beam
was calculated using the beam-beam simulator GUINEA-PIG~\cite{gp}, with the beam parameters taken
from~\cite{schulte,ilcparas}, and the particle ray-tracing was performed using the programme DIMAD~\cite{dimad}. We find that
it is necessary to primarily consider the Low power and High Luminosity beam parameter sets, due to their high beamstrahlung
energy loss. We also conservatively consider the transport of the disrupted beam for collisions with vertical offsets chosen to maximise the corresponding beamstrahlung energy loss. 
Such offsets, typically of about a hundred microns, are expected in the course of tuning the accelerator, in particular during the settling of the trajectory 
feedback system at the IP. For the normal luminosity-producing zero offset collisions, the effect of radiative Bhabha processes is included. 
We use 1.28 million macroparticles to represent the disrupted charged beam. The resulting minimum power loss is of order 0.5 to 1 W, which can hence be considered as the upper bound for cases with zero 
particle loss\footnote{Strictly speaking the 95\% confidence level upper limit is three times this value, or 1.5 to 3 W.}. The radiative Bhabha losses, on the other hand, are not statistics limited.

\subsection{Results for E=500 GeV}

In Table~\ref{tabplbl500} we show, for reference, the power losses for the baseline lattice
at~500~GeV. This lattice is obtained by scaling down the~1~TeV magnet parameters. Table~\ref{tabplnbti500}
shows the power losses for the optimised~500~GeV NbTi lattice, produced following the procedure described in Section~\ref{secopt} and using the limits for the magnet parameters listed in Table~\ref{tabmp}. 
Note that the outer pocket
radius of QF1 was increased by 1mm (from 115mm to 116mm) to accommodate the charged beam of the 
high luminosity parameter set with a vertical offset. Comparison of the performance of the
baseline final doublet at 500 GeV, Table~\ref{tabplbl500}, with the performance of the
new NbTi doublet optimised for 500 GeV, Table~\ref{tabplnbti500}, shows the benefit of 
the new final doublet layout. The power losses for all parameter sets are greatly reduced and, following
the arguments about quenching of superconducting magnets in Section~\ref{secmags}, the power losses 
should be small enough for successful beam transport to be achieved.

\begin{figure*}[tb]
\vspace{0mm} \centering
\includegraphics[width=90mm]{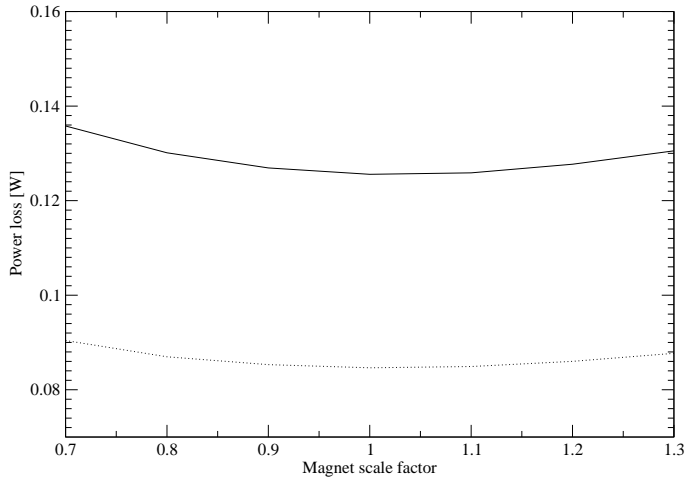}
\vspace{0mm} \caption{The beneficial effect of a 20\% gradient increase, for the High Luminosity parameter
set radiative Bhabha losses. The solid line is the power losses for the NbTi 500~GeV optimised doublet described
in the text, and the dotted line shows the power losses achieved if an extra 20\% of magnetic
gradient became available.}
\label{figpl}
\end{figure*}

Figure~\ref{figpl} illustrates how much the power losses would be further reduced if research and development provided
a stronger (i.e. larger) magnetic gradient. This curve shows the power deposition from radiative Bhabhas for the High Luminosity beam parameter 
set, as a function of magnet scale factor for the optimised NbTi 500 GeV final doublet (solid curve)
and the same magnet with an extra 20\% of gradient in QD0. 

\begin{figure*}[tb]
\vspace{0mm} \centering
\includegraphics[width=90mm]{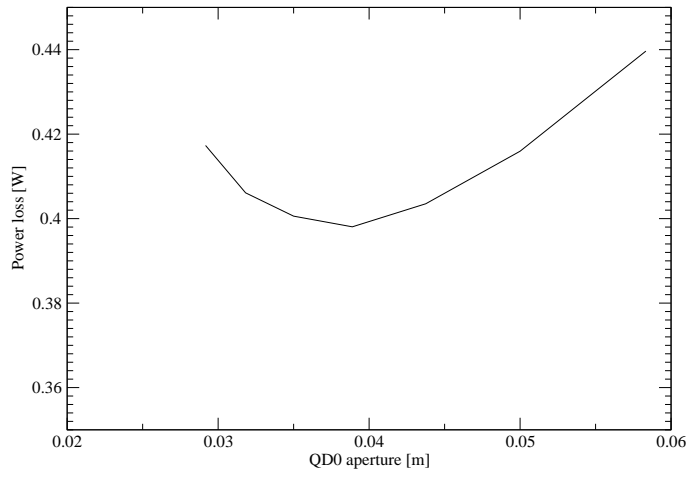}
\vspace{0mm} \caption{The power loss from the Low Power parameter set radiative Bhabhas, as a function
of the QD0 aperture. This plot shows the shallow minimum in the power loss curve, which could be
used to reduce the large external size of the magnet. Note, the
dependence shown here is for the  baseline doublet at 500 GeV. A similarly
shallow behaviour is obtained for the other cases.}
\label{figaper}
\end{figure*}

Figure~\ref{figaper} shows the power deposition from radiative Bhabhas for the Low Power beam parameter set, as a function
of the QD0 aperture\footnote{The
dependence shown here is for the  baseline doublet at 500 GeV. A similarly
shallow behaviour is obtained for the other cases.} . The minimum in the power loss curve is reasonably shallow, which could be
used to reduce the large external size of the magnet. Note that also including the charged
beam losses may move this total minimum to slightly larger aperture.

\begin{figure*}[tb]
\vspace{0mm} \centering
\includegraphics[width=90mm,angle=-90]{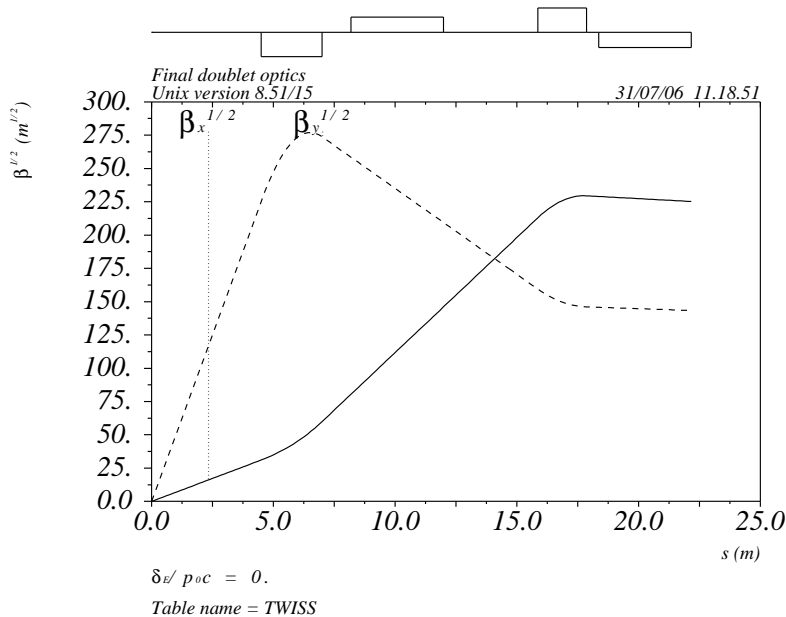}
\vspace{0mm} \caption{The linear optics in the baseline final doublet for the nominal parameters at 500 GeV.}
\label{figopticsbaseline}
\end{figure*}

\begin{figure*}[tb]
\vspace{0mm} \centering
\includegraphics[width=90mm,angle=-90]{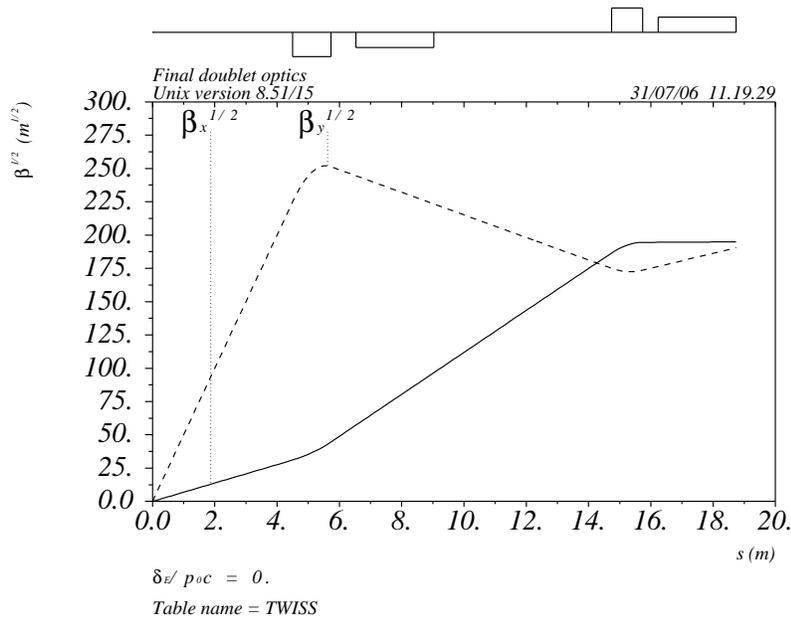}
\vspace{0mm} \caption{The linear optics in the optimised NbTi final doublet for the nominal parameters at 500 GeV.}
\label{figopticsnbti}
\end{figure*}

Finally, the linear optics for the baseline final doublet and the optimised NbTi final doublet at 500 GeV (nominal parameters) can be
seen in figures~\ref{figopticsbaseline} and~\ref{figopticsnbti}.

\begin{table}[htb]
\vspace{-0mm}
\begin{center}
\caption{The power losses on the magnets for the baseline NbTi 500 GeV lattice.}
\vspace{0mm}
\begin{tabular}{|l|c|c|c|c|}
\hline
Beam    		&	QD0	&	SD0	&	QF1	&	SF1  	\\ \hline\hline
Nominal 		&	0 W	&	0 W	&	0 W	&	0 W	\\ \hline
Nominal (dy=200nm)	&	0 W	&	0 W	&	0 W	&	 0 W	\\ \hline
Nomial RB		&	0.40 W	&	0 W	&	0.01 W	&	0.11 W	\\ \hline
low power		&       7.61 W	&	0 W	&	0 W	&     469.76 W	\\ \hline
Low power (dy=120nm)	& 7.77 W	&	0 W	&	0 W	& 188.05 W	\\ \hline	
Low power RB		&	0.39 W	&	0 W	&	0.01 W	& 0.10 W	\\ \hline
High lum.		&    91.81 W	&	0 W	&	0 W	& 1399.82 W		\\ \hline
High lum. (dy=120nm)	& 88.86 W	&	0 W	&       5.42 W	& 192.52 W	\\ \hline
High lum. RB		&	0.97 W	&	0 W	&	0.03 W	& 0.25 W	\\ \hline
\end{tabular}
\label{tabplbl500}
\end{center}
\end{table}

\begin{table}[htb]
\vspace{-0mm}
\begin{center}
\caption{The power losses on the magnets for the optimised NbTi 500 GeV lattice.}
\vspace{0mm}
\begin{tabular}{|l|c|c|c|c|}
\hline
Beam    		&	QD0	&	SD0	&	QF1	&	SF1  	\\ \hline\hline
Nominal 		&	0 W	&	0 W	&	0 W	&	0 W	\\ \hline
Nominal (dy=200nm)	&	0 W	&	0 W	&	0 W	&	0 W 	\\ \hline
Nomial RB		&    0.05 W	&	0 W	& 0.13 W	&	0.03 W	\\ \hline
low power		&	0 W	&	0 W	&	0 W	&	0 W	\\ \hline
Low power (dy=120nm)	&   0 W		&	0 W	&	0 W	&	0 W	\\ \hline	
Low power RB		&	0.05 W	&	0 W	& 0.13 W	&  0.03 W		\\ \hline
High lum.		&	0 W	&	0 W	& 0 W	& 0 W		\\ \hline
High lum. (dy=120nm)	&	0 W	&	0 W	&	0 W	& 0 W		\\ \hline
High lum. RB		&  0.12 W	& 0 W	&  0.33 W	& 0.08 W		\\ \hline
\end{tabular}
\label{tabplnbti500}
\end{center}
\end{table}

\subsection{Results for E=1 TeV}

In Table~\ref{tabplbl1000} we show, for reference, the power losses for the baseline lattice at~1~TeV.
Table~\ref{tabplnbti1000} shows the power losses in the NbTi final doublet, optimised for 1 TeV. The low maximum
gradient possible with this magnet technology means that, whilst the losses are lower than the baseline design
at 1 TeV, they are still too high to avoid magnet quenching. This indicates that higher gradient
magnets are required.

\begin{table}[htb]
\vspace{-0mm}
\begin{center}
\caption{The power losses on the magnets for the baseline 1 Tev lattice.}
\vspace{0mm}
\begin{tabular}{|l|c|c|c|c|}
\hline
Beam    		&	QD0	&	SD0	&	QF1	&	SF1  	\\ \hline\hline
Nominal 		&	0 W	&	0 W	&	0 W	&	0 W	\\ \hline
Nominal (dy=100nm)	&	3.39 W	&	0 W	&	0 W	&	0 W 	\\ \hline
Nominal RB		&	1.02 W	&	0 W	&	0.01 W	&	0.30 W	\\ \hline
Low Power		&	755.05 W &	0 W	&	7.63 W	&	95.37 W	\\ \hline
Low Power (dy=120nm)	&	897.46 W &	0 W	&	20.98 W	&	109.49 W \\ \hline	
Low Power RB		&	1.19 W	&	0 W	&	0.03 W	&	0.34 W	\\ \hline
High Lum.		&	8430.59 W &	0 W	&	190.17 W &	878.12 W \\ \hline
High Lum. (dy=80nm)	&	14076.00 W &	0 W	&	454.53 W &	1514.45 W \\ \hline
High Lum. RB		&	3.60 W	&	0 W	&	0.11 W	&	0.92 W	\\ \hline
\end{tabular}
\label{tabplbl1000}
\end{center}
\end{table}

\begin{table}[htb]
\vspace{-0mm}
\begin{center}
\caption{The power losses on the magnets for the NbTi 1 TeV lattice.}
\vspace{0mm}
\begin{tabular}{|l|c|c|c|c|}
\hline
Beam    		&	QD0	&	SD0	&	QF1	&	SF1  	\\ \hline\hline
Nominal 		&	0 W	&	0 W	&	0 W	&	0 W	\\ \hline
Nominal (dy=100nm)	&	0 W	&	0 W	&	0 W	&	0 W 	\\ \hline
Nominal RB		&	0.73 W	&	0 W	&	0.06 W	&	0.16 W	\\ \hline
Low Power		&      223.94 W &	27.85 W	&	5.76 W	& 	8.07 W	\\ \hline
Low Power (dy=120nm)	&    271.85 W	&	6.77 W	&	11.39 W	&	10.66 W	\\ \hline	
Low Power RB		&   0.15 W	&	0 W	&	0.39 W	&	0.10 W	\\ \hline
High Lum.		&  3169.79 W	&	703.78 W &	54.96 W	&	122.87 W \\ \hline
High Lum. (dy=80nm)	& 5501.98 W	&	409.85 W	&	352.71 W &	236.43 W\\ \hline
High Lum. RB		& 2.55 W	&	0 W	&	0.12 W	&	0.36 W	\\ \hline
\end{tabular}
\label{tabplnbti1000}
\end{center}
\end{table}

\begin{table}[htb]
\vspace{-0mm}
\begin{center}
\caption{The power losses on the magnets for the Nb$_3$Sn 1 TeV lattice.}
\vspace{0mm}
\begin{tabular}{|l|c|c|c|c|}
\hline
Beam    		&	QD0	&	SD0	&	QF1	&	SF1  	\\ \hline\hline
Nominal 		&	0 W	&	0 W	&	0 W	&	0 W	\\ \hline
Nominal (dy=100nm)	&	0 W	&	0 W	&	0 W	&	0 W 	\\ \hline
Nominal RB		&	0.32 W	&	0 W	&	0.34 W	&	0.10 W	\\ \hline
Low Power		&	20.52 W	&	0 W	&	28.73 W	&	27.56 W	\\ \hline
Low Power (dy=120nm)	&	7.77 W	&	0 W	&	48.03 W	&	40.71 W	\\ \hline	
Low Power RB		&	0.37 W	&	0 W	&	0.39 W	&	0.12 W	\\ \hline
High Lum.		&	264.42 W &	0 W	&	477.77 W &	439.98 W \\ \hline
High Lum. (dy=80nm)	&	473.00 W &	0 W	&	1286.15 W &	670.40 W \\ \hline
High Lum. RB		&	1.09 W	&	0 W	&	1.19 W	&	0.36 W	\\ \hline
\end{tabular}
\label{tabplnb3sn1000}
\end{center}
\end{table}

Finally, Table~\ref{tabplnb3sn1000} shows the power losses in the Nb$_3$Sn final doublet, optimised for 1 TeV. 
The higher gradients expected from using Nb$_3$Sn superconductive material enable to significantly reduce the power deposition as compared to both the baseline and optimised NbTi final doublet designs. 
However,
the losses are still in excess of the quenching limits of the magnets. To understand what improvement in superconducting technology would be required, 
we computed how much larger the pole tip field in QD0 would need to be to bring the sum of the charged beam and 
radiative Bhabha losses to less than 1W, for the Low Power beam parameter set.
The result was that a pole tip field of 11T in a magnet with an aperture of 35mm would be needed for the Low Power beam parameter set to be accommodated in the final doublet at 1 TeV. 
The feasibility of such large pole tip fields on the time-scale of the 1 TeV energy upgrade of the ILC needs to be discussed with relevant superconductive magnet experts. Alternatively, beam parameters at 
1 TeV may be optimised with somewhat lower beamstrahlung energy loss, including for total beam powers as low as in the present Low Power parameter set. Such sets would result in significantly easier final 
doublet designs. 

\section{CONCLUSIONS}
\label{secconc}
In this note, we have described a detailed study of the power deposition in the final doublet of the ILC 2mrad crossing-angle layout, from both the disrupted 
charged beam and from the radiative Bhabha events. We exploited superconductive magnet technologies both currently available and expected to be developed through on-going R\&D: NbTi and 
Nb$_3$Sn with maximum pole tip fields of respectively 6.28 T and 8.8 T.

We find that using NbTi technology to optimise the 500 GeV baseline final doublet is adequate to obtain small enough beam power depositions in the final doublet, 
provided a Tungsten liner is included in the design of the vacuum chamber to spread out the showers and reduce the maximum power density at the location where most losses are concentrated. 
This doublet now needs to be integrated into the final focus system and extraction line. 
When this will be done, we will propose that it become the baseline for the 2mrad at 500GeV, with the assumption that it would have to 
be replaced at the time of the 1 TeV energy upgrade. In all cases, the sextupole lengths, strengths and apertures are
not fully optimised. Some more work is still needed here, in view of reducing their sizes and ease their integration with the detector.
It is also necessary to include the extraction line magnets in a further optimisation.

A final doublet based on NbTi technology does not, on the other hand, provide small enough beam power depositions at 1 TeV, for several of the beam parameter sets considered. 
Exploiting Nb$_3$Sn superconductive technology can however improve the situation considerably, as the higher expected gradients would allow significantly lower power depositions.
The optimised layout shows far superior power loss behavior than the current
1 TeV machine baseline final doublet, but is still in excess of the quenching limit for the beam parameter sets with the 
largest beamstrahlung energy loss. A 35mm QD0 with a pole tip field
of around 11 T would be needed to a avoid quenching of this magnet, in the case of the Low Power beam parameters.

We expect further R\&D in superconductive magnet technology to enable increases in maximum available pole tip field in the future, which will ease the design of the 2mrad beam crossing-angle layout. An alternative path is to constrain the optimisation of the beam parameters at 1 TeV with an upper bound on the beamstrahlung energy loss. This would also be motivated since a very large beamstrahlung energy loss is undesirable for several other reasons. 

\section*{Acknowledgement}
We would like to thank Andrei Seryi and Deepa Angal-Kalinin for helpful advice and assistance, Grahame Blair for useful discussions and John Carter for his assistance computing the effect of a Tungsten liner in the vacuum chamber.

\end{document}